\documentclass[english,sort&compress]{article}
\usepackage{geometry}
\usepackage[T1]{fontenc}
\usepackage[latin9]{inputenc}
\usepackage{array}
\usepackage{longtable}
\usepackage{booktabs}
\usepackage{units}
\usepackage{amsmath}
\usepackage{graphicx}
\usepackage{esint}

\usepackage{authblk}

\makeatletter

\providecommand{\tabularnewline}{\\}

\usepackage{cite}

\makeatother

\usepackage{babel}
\begin{document}
\global\long\def\d{\text{d}}
\global\long\def\d{\text{d}}
\global\long\def\red#1{\tilde{#1}}
\global\long\def\unit#1{\,\mathrm{#1}}

\global\long\def\Length{\mathcal{L}}

\global\long\def\Energy{\mathcal{E}}

\global\long\def\zmax{\hat{z}}

\global\long\def\ArcLength{\mathcal{S}}
\global\long\def\SurfaceArea{\mathcal{\mathcal{A}}}
\global\long\def\Volume{\mathcal{V}}
\global\long\def\Circumference{\mathcal{C}}

\global\long\def\ParameterForHead#1{#1_{\mathrm{head}}}
\global\long\def\ParameterForNeck#1{#1_{\mathrm{neck}}}
\global\long\def\ParameterForBase#1{#1_{\mathrm{base}}}
\global\long\def\ParameterForFilopodium#1{#1_{\mathrm{filop.}}}
\global\long\def\ParameterForActin#1{#1_{\mathrm{actin}}}

\global\long\def\fHead{\ParameterForHead f}
\global\long\def\fNeck{\ParameterForNeck f}
\global\long\def\fActin{f_{\mathrm{actin}}}

\global\long\def\RHead{\ParameterForHead R}
\global\long\def\RNeck{\ParameterForNeck R}
\global\long\def\RBase{\ParameterForBase R}
\global\long\def\LNeck{\ParameterForNeck{\Length}}
\global\long\def\LFilopodium{\ParameterForFilopodium{\Length}}

\global\long\def\SurfaceAreaHead{\ParameterForHead{\SurfaceArea}}
\global\long\def\SurfaceAreaNeck{\ParameterForNeck{\SurfaceArea}}
\global\long\def\SurfaceAreaFilopodium{\ParameterForFilopodium{\SurfaceArea}}

\title{Actin Remodeling and Polymerization Forces Control Dendritic Spine
Morphology }

\author[1]{Karsten Miermans\footnote{Corresponding author, k.miermans@gmail.com}}
\author[1]{Remy Kusters}
\author[3]{Casper Hoogenraad}
\author[1,2]{Cornelis Storm}
\affil[1]{Theory of Polymers and Soft Matter, Department of Applied Physics, Eindhoven University of Technology}
\affil[2]{Institute for Complex Molecular Systems, Eindhoven University of Technology}
\affil[3]{Cell Biology, Faculty of Science, Utrecht University}


\maketitle


\begin{abstract}

Dendritic spines are small membranous structures that protrude from the neuronal dendrite. Each spine contains a synaptic contact site that may connect its parent dendrite to the axons of neighboring neurons. Dendritic spines are markedly distinct in shape and size, and certain types of stimulation prompt spines to evolve, in fairly predictable fashion, from thin nascent morphologies to the mushroom-like shapes associated with mature spines. This striking progression is coincident with the (re)configuration of the neuronal network during early development, learning and memory formation, and has been conjectured to be part of the machinery that encodes these processes at the scale of individual neuronal connections. It is well established that the structural plasticity of spines is strongly dependent upon the actin cytoskeleton inside the spine. A general framework that details the precise role of actin in directing the transitions between the various spine shapes is lacking. We address this issue, and present a quantitative, model-based scenario for spine plasticity validated using realistic and physiologically relevant parameters. Our model points to a crucial role for the actin cytoskeleton. In the early stages of spine formation, the interplay between the elastic properties of the spine membrane and the protrusive forces generated in the actin cytoskeleton propels the incipient  spine. In the maturation stage, actin remodeling in the form of the combined dynamics of branched and bundled actin is required to form mature, mushroom-like spines. Our model identifies additional factors that plausibly aid the stabilization and maintenance of spine morphology. Taken together, our model provides unique insights into the fundamental role of actin remodeling and polymerization forces during spine formation and maturation.
\end{abstract}\


\section{Introduction}

A single neuron can contain hundreds to thousands of dendritic spines, actin-rich, micron-sized protrusions
which project from dendritic shafts~\cite{Hotulainen2010}. Mature spines
consist of two basic compartments:~a constricted region called the {\em neck}, supporting a bulbous
{\em head} containing the postsynaptic site that makes contact with the
axon of a nearby neuron. Spines come in a wide range of sizes and shapes,
their lengths varying between $0.2-2\unit{\mu m}$ and their volumes between $0.001-1\unit{\mu m^{3}}$.
Electron microscopy (EM) studies have identified several morphological
categories of spines, such as thin, filopodium-like protrusions (\textquoteleft thin
spines\textquoteright ), and spines with a large bulbous head (\textquoteleft mushroom
spines\textquoteright )~\cite{Harris1988,Harris1989,Harris1992,Hotulainen2010,Korobova2010}.
Different live cell-imaging techniques have demonstrated that dendritic
spines are highly dynamic structures, subject to constant morphological change even after birth.

\begin{center}
\begin{longtable}{>{\centering}p{0.06\columnwidth}>{\raggedright}p{0.45\textwidth}>{\centering}p{0.27\textwidth}>{\centering}p{0.07\textwidth}}
\caption{Various `ball-park' figures of dendritic spines.\label{tab:ballpark-figures}}
\tabularnewline
\endfirsthead
\midrule 
\textbf{Code} & \textbf{Quantity} & \textbf{Typical Scale} & \textbf{Source}\tabularnewline
\midrule
$N$ & Number of actin filaments in spine-head (see supporting information
on how this number was estimated) & $\sim71$ & \cite{Korobova2010}\tabularnewline
$\RBase$ & Radius of the base of the spine (viz. where the spine is connected
to the dendritic membrane). This quantity was estimated on the basis
of microscopy images published by~\cite{Toennesen2014}. & $\sim300\,\mathrm{nm}$ & \cite{Toennesen2014}\tabularnewline
$\RNeck$ & Radius of a typical spine-neck & $75\pm30\,\unit{nm}$ & \cite{Harris1989}\tabularnewline
$\RHead$ & Radius of a typical spine-head & $220\pm154\,\unit{nm}$ & \cite{Harris1989}\tabularnewline
$\LNeck$ & Length of a typical spine-neck & $0.2-2\,\unit{\mu m}$ & \cite{Toennesen2014}\tabularnewline
$\LFilopodium$ & Length of a typical filopodium & $0.9-10\,\mathrm{\mu m}$~\cite{Evers2006} (mean $\approx5\,\mathrm{\mu m}$~\cite{Ziv1996}) & \cite{Evers2006,Ziv1996}\tabularnewline
$\ell$ & Length that actin filament extends upon one polymerization step & $2.2\,\unit{nm}$ & \cite{Mogilner2003}\tabularnewline
$\SurfaceAreaNeck$ & Surface-area of a typical spine-neck & $0.24\pm0.17\,\unit{\mu m^{2}}$ & \cite{Harris1989}\tabularnewline
$\SurfaceAreaHead$ & Surface-area of a typical spine-head & $0.61\pm0.57\,\unit{\mu m^{2}}$ & \cite{Harris1989}\tabularnewline
$\SurfaceAreaFilopodium$ & Surface-area of a typical filopodium (this was calculated using $\SurfaceAreaFilopodium\approx2\pi R_{\mathrm{filop.}}\LFilopodium$
with $\ParameterForFilopodium R,\LFilopodium$ from~\cite{Evers2006,Ziv1996}) & $0.85-16\,\mathrm{\mu m^{2}}$ (mean $\approx6.3\,\mathrm{\mu m^{2}}$) & \cite{Evers2006,Ziv1996}\tabularnewline
$\fNeck$ & Expansive force that a spine-neck of typical size exerts (using the
approximate formula $\fNeck\approx4\pi K_{b}\RHead/\RNeck^{2}$, see
text) & $9-290\,\unit{pN}$ & \tabularnewline
$\fHead$ & Contractile force that a spine-head of typical size exerts (using
the approximate formula $\fHead\approx17.23K_{b}\RNeck/\SurfaceAreaHead$,
see text) & $0.0007-0.05\,\unit{pN}$ & \tabularnewline
$\fActin$ & Average actin polymerization force & $3.8\,\unit{pN}$ & \cite{Mogilner2003}\tabularnewline
$K_{b}$ & Bending rigidity of lipid bilayer membrane & $5\times10^{-19}\,\unit{J}$ & \cite{Semrau2008}\tabularnewline
\bottomrule
\end{longtable}
\par\end{center}

\begin{figure}
\centering\includegraphics{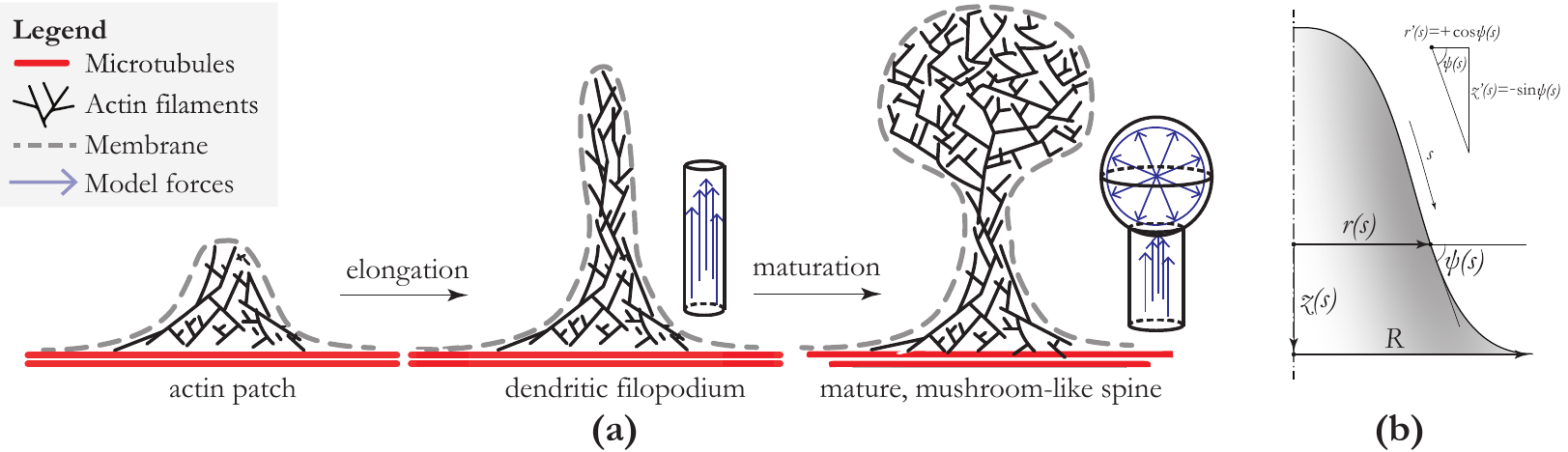}

\caption{Outline of the model for spine formation and maturation. Panel (a):
Cartoon of spine initiation, elongation and maturation. From left
to right: \textquoteleft stubby spines\textquoteright ; dendritic
filopodia or thin spines; mature, mushroom-like spines. In our mathematical
model, we solve the shape equation based on the energy functional
(1) (see SI, equation (3)). In this study we show that, at least for
the purposes of the force calculations, the results of the shape equation
can be reproduced using the geometries that are also displayed in
this figure. Panel (b): Definition of axisymmetric coordinate system
that use for our models. \label{fig:model-outline}}
\end{figure}

During neuronal development, dendrites initially appear as thin and hairlike
filopodia (figure~\ref{fig:model-outline}). They are defined as
having a length that is at least twice the width, and they do not
display the bulbous head found on dendritic spines~\cite{Korobova2010,Yasumatsu2008,Kasai2010}.
Filopodia are devoid of organelles and vesicles, and are composed
primarily of actin filaments. These actin filaments are bundled and primarily aligned to the nascent spine. Filopodia are the precursors to dendritic spines, and their flexibility allows the establishment of synaptic contacts. Once the contact between a dendritic filopodium and a neighboring axon has been established, the spine-head begins to swell, taking on a
more mushroom-like morphology. Over time, such recognizable mushroom spines become the prevalent structure on the dendritic shaft, and few filopodia remain.

The progressive shape change is neither random nor deterministic. Rather, it is  thought to be correlated with the strength and maturity of each synapse~\cite{Yasumatsu2008,Kasai2010}.
At the level of an individual spine, strengthening of a synapse is
accompanied by modifications in the size of the spine. The prime mechanisms
that drives structural plasticity is the modulation of actin dynamics
in dendritic spine. Although the importance of actin remodeling as
well as the synaptic signaling mechanisms involved in structural synaptic
plasticity are well established~\cite{Korobova2010,Hotulainen2010,Okamoto2004},
a general framework to correlate the state of the actin cytoskeleton to spine shape is lacking. Most importantly, it is not clear whether the actin is capable of autonomously {\em driving} the shape change, or whether the actin simply {\em follows} morphological transitions otherwise imposed.

Our model for spine dynamics uses the Canham-Helfrich
formalism, an approach which has proven its strength in describing,
both qualitatively and quantitatively, the deformation of biological
membranes in numerous biological systems such as red-blood cells~\cite{Angelov2000},
membrane tethers~\cite{Derenyi2002} and binary or tertiary lipid
mixtures in giant-unilamellar vesicles~\cite{Semrau2008}. For a broad overview, we refer to \cite{Seifert1997} and many references therein. We analyze
the interplay of the plasma membrane with the underlying actin cytoskeleton
to quantify the forces that are required to prompt the initial formation of the spine, and its subsequent outward growth. We find that the forces generated by actin polymerization are sufficient for it to drive filopodium formation, and that the resulting dimensioning (quantified, for instance, by the ratio $\mathrm{(protrusion\,width)/length}$) closely resembles those reported in experiments. A related theoretical model taking into account the  interplay of the spine membrane with the actin cytoskeleton allows us, in addition, to compute the forces and energies required for spine head formation. It shows that the simultaneous presence of both branched actin filaments and bundled/aligned actin is required, and sufficient, to produce the typical mushroom-like spine morphology. Finally, our model also highlights the important role of additional physical processes in stabilizing the morphological features of mature spines. We discuss several candidate factors that may effect these processes, and conclude that these molecules are sufficiently rigid to be able to constrict the spine-neck to the extent reported in experiments. Our models do point to a fundamental role for actin remodeling in the process of spine formation and maturation. This finding supports earlier claims in the literature, and  our model suggests novel experiments to further pin down the basic principles that control the structural plasticity of the brain.

\section{Materials and Methods}

Reflecting the approximate rotational symmetry of dendritic spines,
we use an axisymmetric coordinate system consisting of an angle $\psi$
with the horizontal, an arc-length parameter $s$, radial coordinate
$r$ and vertical coordinate $z$. The coordinate system is schematically
displayed in figure \ref{fig:model-outline}. The arc-length parameter
$s=0...\ArcLength$ is used as the independent variable and $r(s)$
and $\psi(s)$ as the coordinates. This coordinate system fully determines
the shape, and the vertical coordinate $z(s)$ is recovered by the
geometrical relation $z'(s)=-\sin\psi(s)$. The Canham-Helfrich energy
functional that we use can be written \cite{Derenyi2002,Semrau2008}
\begin{equation}
\mathcal{F}=\tfrac{1}{2}K_{b}\int\d a\,(2H)^{2}+\sigma(\SurfaceArea-\SurfaceArea_{0})-f(\Length-\Length_{0}),\label{eq:CH-energy-functional}
\end{equation}
where $K_{b}\approx500\unit{pN\cdot nm}$ is the bending rigidity
of the membrane~\cite{Semrau2008}, $2H=\psi'(s)+\sin\psi(s)/r(s)$
is the mean curvature~\cite{Juelicher1994} (with $\psi'(s)\equiv\d\psi/\d s$),$\sigma$
is a surface tension,\footnote{This surface tension is a parameter that is used to enforce the total
surface-area (a ``Lagrange multiplier''), and can therefore not be
interpreted (even though it has the same dimensions) as the surface
tension of~$\approx0.05\unit{pN/nm}$ that is measured in experiments
using e.g.~membrane tethers~\cite{Raucher1999,Gauthier2012,Diz-Munoz2013}.\label{fn:surface-tension-explanation}} $\SurfaceArea=\int\d a$ is the surface area, $f$ is a point-force
acting on the membrane and $\Length=z(\ArcLength)-z(0)$ is the height
of the membrane. The first term in this energy functional---the one
containing the mean curvature $2H$---represents the bending energy
of the membrane, which reflects the tendency of lipid bilayers to
adopt a flat shape (or spherical in the case of vesicles). We use
the surface tension~$\sigma$ and point-force~$f$ as Lagrange multipliers
to enforce specific values of the surface-area~$\SurfaceArea_{0}$
and the height of the shape~$\Length_{0}$~\cite{MichaelStone2009}.
Within this paradigm, we interpret the surface-area, viz.~amount
of membrane available to the spine, as a quantity that encodes growth~\cite{Park2006,Wang2008}.
The height of the shape reflects the cytoskeletal architecture of
the spine. Although there is no obvious way of interpreting the surface
tension (cf.~footnote~\ref{fn:surface-tension-explanation}), the
point-force~$f$ is simply the mechanical force that is exerted by
the cytoskeleton on the spine membrane. One of the main goals of this
paper is to investigate whether these forces are attainable through
actin polymerization. We will show that this is indeed the case.

Our choice to work at fixed total surface-area, rather than fixed
surface tension, is inspired by two considerations: (i)~Dendrites
are finite in size, and the membrane that envelopes the dendritic
shaft can only be as small as the underlying cytoskeleton of microtubules~\cite{Korobova2010,Hotulainen2010}---therefore,
we cannot regard the surroundings of the spine as a reservoir of freely
accessible membrane. Instead, excess membrane needs to be transported,
often by means of exocytic trafficking, in order to be available to
the spine~\cite{Park2006,Wang2008}. (ii)~On the dendritic shaft,
generally, many spines exist side-by-side. In open boundary settings,
such as those employed in \cite{Derenyi2002}, area is exchanged with
a virtual bath outside the integration domain. In the dendritic shaft,
however, no such bath exists as the next spine is likely also growing.
Thus, a competition for membrane exists between proximate spines.
For this reason, we choose to work with closed boundaries, prohibiting
area to leak out of the domain of interest.

We choose to work in a setting in which the amount of membrane available
to the spine is conserved, i.e. membrane does not leak away from the
shape. We point out that there is biological evidence that cells strive
to maintain their surface-tension~\cite{Gauthier2012,Raucher1999,Derenyi2002}.
Although this empirical fact might seem incompatible with our simulations,
the ensembles of constant surface-area and constant surface-tension
are---for the purpose of modeling mushroom-like spines---approximately
equivalent. The underlying reason for this similarity is simple: the
energy corresponding to the surface-tension of a typical spine-head
is approximately equal to the bending energy of a typical spine-neck.
Hence, for mushroom-like shapes, the two ensembles can be converted
to one another by swapping an energy term of the same order of magnitude.
We have verified this prediction by comparing various quantitative
predictions between the two ensembles.\footnote{We have compared predictions of the filopodium width, the number of
filaments required to support a filopodium, the neck width of a mature
spine and the head width. Of these metrics, the mean values of the
predictions in these two ensemble differ $5-40\%$.} Indeed, the predictions that we make in this paper are valid in both
ensembles.

Using the Euler-Lagrange formalism, the energy functional (\ref{eq:CH-energy-functional})
can be transformed into a system of differential equations. These
shape equations, reproduced in the Supporting Information, have been
numerically solved by means of a shooting-and-matching technique for
a wide range of parameters $\SurfaceArea_{0},\Length_{0}$ and several
sets of boundary conditions (we drop the subscripts to $\SurfaceArea_{0},\Length_{0}$
in the remainder of the paper). We ignore the stretching energy since
lipid bilayer membranes can be regarded as approximately inextensible~\cite{Rawicz2000}.
Also, the pressure-term $-p\Volume$ that is often cited in conjunction
with the Canham-Helfrich formalism \cite{Seifert1997,Juelicher1994,Juelicher1996}
is not applicable since we are discussing a system that is free to
exchange cytosol with the environment.

\section{Results and Discussion}

We use the Canham-Helfrich energy functional (\ref{eq:CH-energy-functional})
to model the growth of dendritic spine membranes. The growth sequence
is schematically shown in figure~\ref{fig:model-outline}. We will
show that this growth sequence can be explained
qualitatively and quantitatively by simple models that incorporate
the interaction of the actin cytoskeleton and the spine membrane.
To that end, we will first determine how filopodia are formed by application
of forces that the cytoskeleton exerts on the spine membrane. Then,
we will show that the forces generated by a branched cytoskeleton,
located at the top of the spine, will result in a bulbous head and
a thin spine-neck. Finally, we will show that actin-membrane anchoring
or ring-like molecules are another scenario for constraining a large
head and long, thin neck. For the model calculations, we shall make
repeated use of the physiologically relevant parameters that we have
tabulated (see table~\ref{tab:ballpark-figures}, Supporting Information).

\subsection{Filopodium Formation}

\begin{figure}
\centering\includegraphics{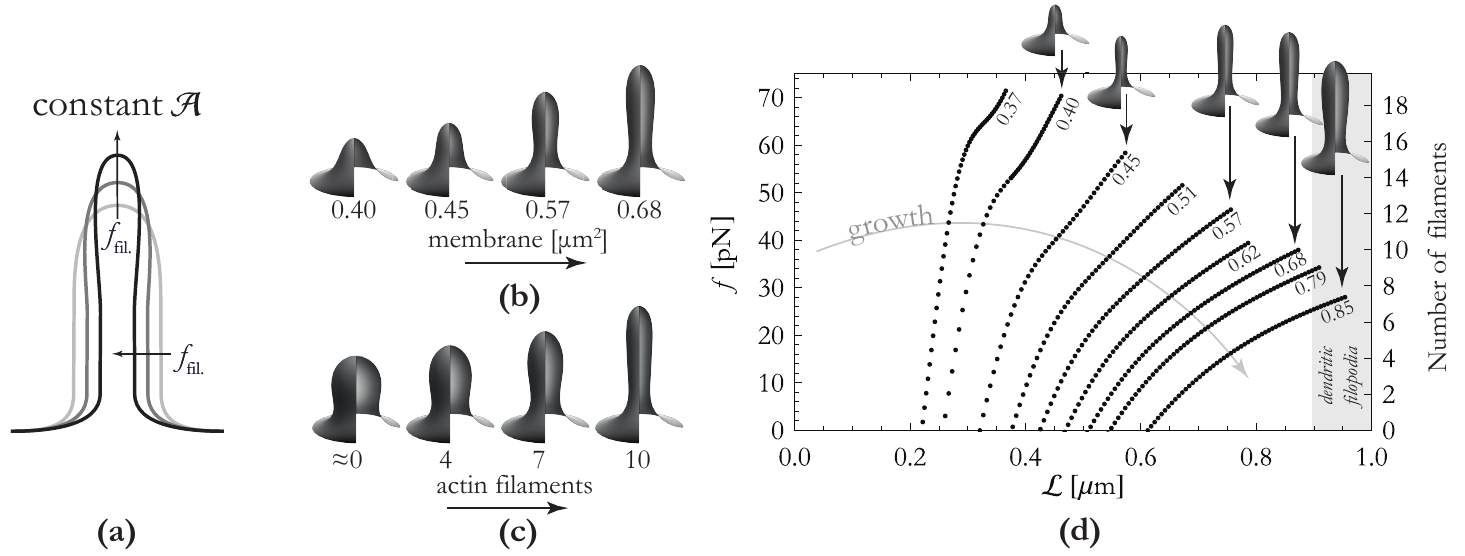}

\caption{Outline of results for filopodium formation. Panel (a): Cartoon of
qualitative effect of increasing force whilst the amount of membrane
is kept constant. Panel (b): Effect of growth (viz. membrane addition)
on filopodium morphology (if the force on the membrane is kept constant).
These shapes experience a vertical force of $35\,\mathrm{pN}$ corresponding
to approximately 9 polymerizing actin filaments. Membrane addition
results in substantial elongation of the filopodium. Panel (c): Effect
of cytoskeletal remodelling (viz. actin polymerization) on filopodium
morphology if the amount of membrane is kept constant. These shapes
have a surface-area of $0.68\,\mathrm{\mu m^{2}}$. Increasing the
number of polymerizing actin filaments leads to a marked change in
morphology from a stubby-like morphology to a tubular shape..\protect \\
Panel (d): Force-extension curves of our models of dendritic filopodia
for various values of the surface-area~$\protect\SurfaceArea$. Numbers
at curves indicate the surface-area in units of $\mathrm{\mu m^{2}}$
whereby we used a radius of the base of the filopodia $\protect\RBase=300\,\unit{nm}$.
\label{fig:force-extension-curves-with-shapes}}
\end{figure}

It is well known that the actin cytoskeleton plays a large role in
the formation of filopodia \cite{Hotulainen2010}. It has been hypothesized
that polymerization of actin filaments and the resultant forces are
sufficient for the formation of dendritic filopodia \cite{Borisy2000}.
In order to theoretically investigate this possibility, we will present
a model that includes extension of the actin cytoskeleton in growing
filopodia. This is schematically displayed in the outline of our model,
figure \ref{fig:force-extension-curves-with-shapes}. This is readily
incorporated in the energy functional (\ref{eq:CH-energy-functional})
by fixing the height of the shape---thereby representing the vertical
dimension of the cytoskeleton. Growth of the cytoskeleton, or change
in cytoskeletal architecture, is represented by incrementing this
height constraint. We will first show that the forces that this rigid
structure needs to exert on the spine membrane match the forces that
are generated by actin polymerization. Then, we will show that the
sequence of shapes as a consequence of polymerization of the actin
cytoskeleton is similar to that of filopodium formation.

The protrusive forces that the rigid actin cytoskeleton exerts on
the spine-membrane, will result in tube-like shapes, as can be seen
in figure \ref{fig:force-extension-curves-with-shapes}. Also shown
are the force-extension curves of these tubes for various values of
the membrane surface-area. Since growth of dendritic spines and filopodia
is mediated by exocytosis of endosomes at the synapse \cite{Park2006,Wang2008},
we can model the growth of spines by increasing the surface-area of
the shape (also see the Methods part of this paper). Thus, we find
that filopodia with more membrane require less force to be extended---in
other words, membrane addition will result in further elongation of
filopodia. Although the full force-extension relation shown in figure~\ref{fig:force-extension-curves-with-shapes}
is non-trivial, the linear part for large extensions (i.e. large height
$\Length$) can easily be understood from a theory that treats these
structures as cylinders. From the bending energy of a cylinder (with
given surface-area) $\Energy=2\pi^{2}K_{b}\Length^{2}/\SurfaceArea$
we find that the force~$f\equiv-\partial_{\Length}\Energy$ to extend
this cylinder is linear in $\Length$. Applying this derivative, we
find the force for producing filopodia $f\approx4\pi^{2}K_{b}\Length/\SurfaceArea$.
This approximation turns out to be accurate to within 9\% of the computed
force-extension curves shown in figure \ref{fig:force-extension-curves-with-shapes}
(the error in this approximation decreases as the filopodium height
increases). This is markedly different from the force required for
pulling a tube from a reservoir (i.e. a (quasi-)infinite bath of membrane)
of surface area. As is discussed in \cite{Derenyi2002} (using detailed
analytical and numerical calculations) the force for pulling a tube
from such a reservoir converges to a constant for large extensions.
If it were the case, then, that dendritic filopodia were connected
to a bath of membrane, we would not expect dendritic filopodia to
have a typical length. On the contrary, in that scenario dendritic
filopodia would grow \emph{ad infinitum} (given that the applied force
is large enough to overcome an initial barrier). We assert that, within
the paradigm of a conserved quantity of membrane available to the
spine, a finite force will result in a definite length of the filopodia.

As can be seen in figure~\ref{fig:force-extension-curves-with-shapes},
the force required for formation of dendritic filopodia is in the
tens of piconewtons. The polymerization of actin is able to exert,
on the average, a force of $\fActin\approx3.8\,\mathrm{pN}$ (from~\cite{Mogilner2003},
see Supplementary Information). We find, using the typical values
for the length and surface-area (table~\ref{tab:ballpark-figures},
see Supporting Information) and the aforementioned formula $f\approx4\pi^{2}K_{b}\Length/\SurfaceArea$,
a minimal number of actin filaments of~$5-15$. Although we have
not been able to find publications that mention the number of actin
filaments in dendritic filopodia, examining EM of the cytoskeletal
organization of dendritic filopodia from~\cite{Korobova2010} suggests
that filopodia typically have~$6-10$ filaments. This comparison
tentatively verifies the plausibility of our model for filopodium
formation.

Our simulations span up to $\SurfaceArea=0.85\,\mathrm{\mu m^{2}}$
and $\Length\approx950\,\mathrm{nm}$. Following~\cite{Ziv1996,Evers2006},
these shapes can be regarded as relatively small filopodia. Now, measuring
the width of the corresponding tubular part of the shape, we find
diameters in the order of $160-200\,\mathrm{nm}$. Indeed,~\cite{Harris1989}
report values for the diameters of filopodia or thin spines in the
range~$90-210\,\mathrm{nm}$. Thus it is found that the simulations
and experimental results have compatible ranges.

\subsection{The Role of the Actin Cytoskeleton in Spine Maturation}

\begin{figure}
\centering\includegraphics[width=0.95\textwidth]{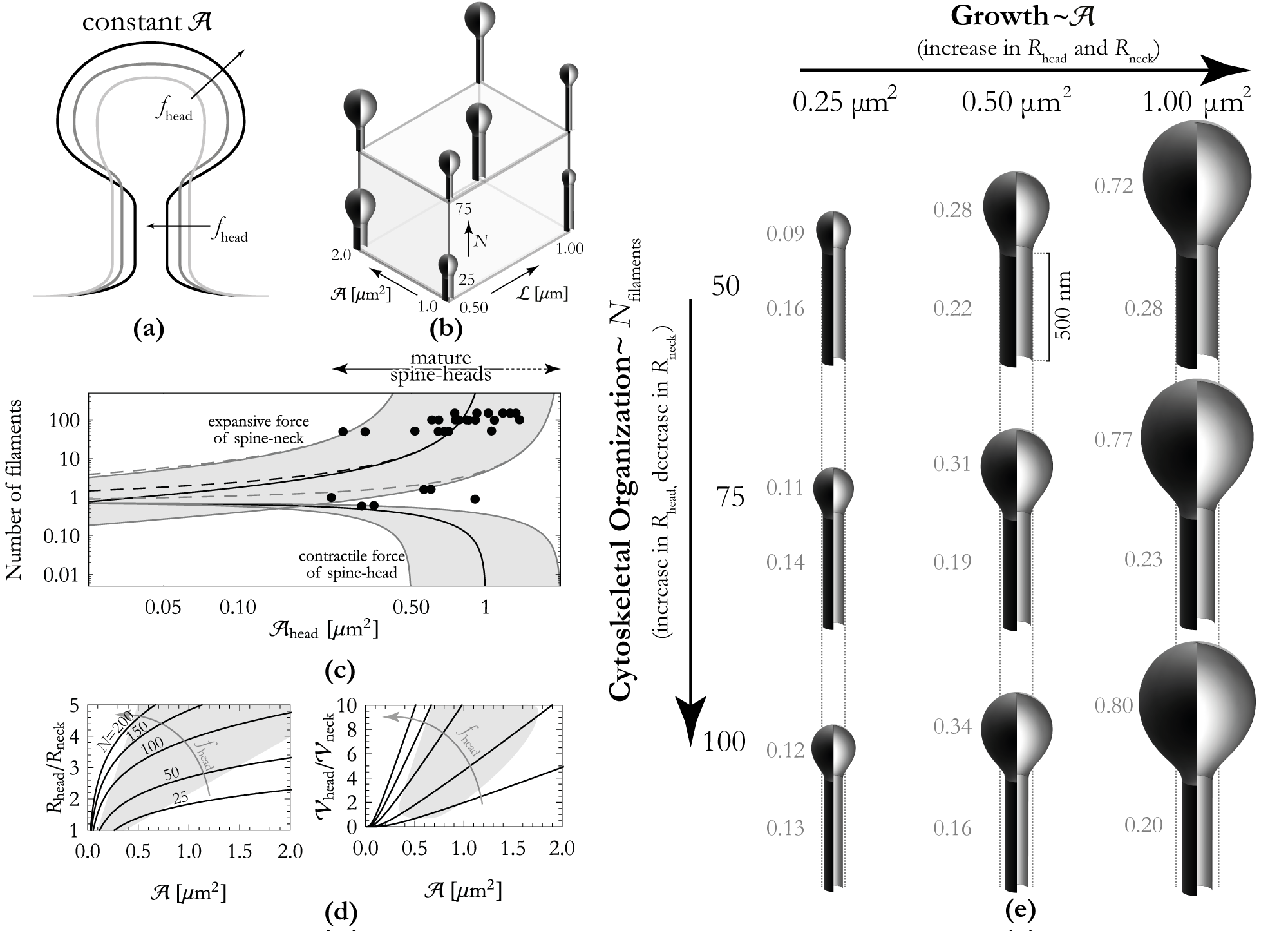}

\caption{Outline of results for spine maturation.\textbf{ }The spine bases
have been left out in the renders in this figure. Panel (a): Cartoon
showing qualitative effect of increasing force (viz.~increasing the
number of filaments in the spine-head) whilst the amount of membrane
is kept constant. Increasing the number of actin filaments in the
spine-head enlarges the spine-head and, at the same time, a thinning
neck. Panel (b): Results of our model are combined in a three-dimensional
growth-organization matrix, shown here with selected shapes. These
shapes show clearly the effects of increasing the number of filaments
in the head $N$, the total surface-area~$\protect\SurfaceArea$
and the length of the spine-neck $\protect\Length$. Panel (c): The
minimum number of actin filaments required in the cytoskeleton for
sustaining the contractile force $\protect\fHead$ that the spine-head
membrane exerts and for counteracting the expansive force $\protect\fNeck$
of the spine-neck. Band indicates typical values of the total amount
of membrane~$\protect\SurfaceArea=0.5\ldots2.0\,\mathrm{\mu m^{2}}$
(cf. table~\ref{tab:ballpark-figures}). Dashed lines indicate number
of actin filaments required for counteracting $\protect\fHead+\protect\fNeck$.
Empirical data (black circles) shows reasonable agreement with our
model (data taken from from~\cite{Frost2010}, see SI). Panel (d):
Ratio of head and neck radii (left) and volumes (right) for a number
of actin filaments $N=25,50,100,150,200$ (lower to upper curves).
For these plots we used equation~(\ref{eq:spine-maturation-implicit-equation-for-spine-head})
with~$\protect\LNeck=500\unit{nm}$. Experimental data from~\cite{Toennesen2014}
is highlighted in gray. Panel (e): Effects of growth (membrane addition)
and the number of actin filaments in the spine-head on spine morphology.
In these models, we kept the total length of the spine-neck fixed.
Dotted lines are a visual aid for showing how increasing the number
of actin filaments in the spine-head results decreases the width of
the spine-neck.\label{fig:spine-maturation-panel}}
\end{figure}

As a consequence of synaptic activity, the spine volume may increase
and there is a marked change in the qualitative morphology through
formation of a bulbous spine-head~\cite{Yang2009,Engert1999}. We
have previously shown that simply adding membrane to dendritic filopodia
results in larger filopodia, but not formation of a bulbous head.
Therefore, an additional process is needed in order to produce mature
spines. By which mechanisms does this qualitative change in morphology
occur? In this part of the paper, we will show that the process of
spine maturation can, at least in part, be ascribed to the interaction
of the spine membrane with an isotropic actin meshwork.

As is the case for filopodia, it is known that the actin cytoskeleton
is intimately linked to the size and shape of the spine head~\cite{Hotulainen2010,Okamoto2004},
and therefore is essential to understanding spine maturation. By modeling
the interaction of the cytoskeleton with the spine membrane, we will
investigate the mechanical requirements for a volume increase and
morphological transition (that is characteristic of spine maturation)
to occur. We will show (by making use of the Canham-Helfrich energy
(\ref{eq:CH-energy-functional}) and comparison with experiments)
that branched actin filaments in the spine-head are plausibly responsible
for the transition from dendritic filopodium to a mushroom-type morphology.
Although the models in this paper lack many of the biological details
relevant for spine maturation, we find that the forces that are required
for spine membrane match the forces that are generated by actin polymerization.
Then, we show that typical spine-neck widths are accurately predicted
by our models.

An outline of our model for spine maturation is displayed in figure
\ref{fig:spine-maturation-panel}. We model the polymerization of
actin in the spine-neck as a vertical force and polymerization in
the spine-head as a radial force. The discrepancy between these two
types of forces stems from the difference in cytoskeletal organization
in spines---the spine-head predominantly contains branched actin whereas
oriented or linear actin mainly localizes in the spine-neck \cite{Hotulainen2010,Korobova2010}.
This gives rise to an approximately isotropic network of actin in
the spine-head, contrary to the actin organization in spine-necks
and dendritic filopodia \cite{Korobova2010}. The polymerization of
these two manifestations of actin result respectively in a radial
force and a directed force. As is the case for our models for dendritic
filopodia, we approximate the total surface-area of the spine as a
constant since there is only a finite pool of membrane available on
the dendrite (for a more detailed explanation, see the Methods part
of this paper). This approximation, combined with the fact that lipid
membranes are practically inextensible \cite{Rawicz2000}, leads to
the following assertion: exerting an outward force on the spine-head
results in the transportation of membrane from the spine-neck to the
spine-head. More simply stated, cytoskeletal growth in the spine-head
results in an increase of the size of the spine-head at the expense
of a decrease in the neck width.

In order to make the above considerations quantitative, we propose
a model for the spine membrane that is composed of a cylinder of constant
height (but variable radius) connected to a sphere. As displayed in
figure \ref{fig:spine-maturation-panel}, we model $N$ filaments
in the spine-head that each apply an outward radial force $\fActin=\fHead/N$
over a radius $\RHead$. The work performed by this force is $\fHead\RHead$.
Likewise, the energy required for attaining a neck of length~$\LNeck$
and radius~$\RNeck$ is~$\pi K_{b}\LNeck/\RNeck$~(equation~(\ref{eq:CH-energy-functional})).
Balance of forces dictates that the total energy~$\pi K_{b}\LNeck/\RNeck-\fHead\RHead$
is minimized\emph{.}\footnote{We have shown by means of theoretical modeling and numerical simulations
that we can neglect the force that is exerted by the contractile force
due to the bending energy of the head. This fact is reflected in the
small number of actin filaments required to sustain a large, bulbous
head, as can be seen in figure~\ref{fig:spine-maturation-panel}(b).} In order to insist conservation of membrane, we insert into the balance
of forces the equation~$\SurfaceArea\approx2\pi\RNeck\LNeck+\SurfaceAreaHead$
with $\SurfaceArea$ a constant. Taken together, this results in the
following implicit equation that we can solve for~$\SurfaceAreaHead$:\nopagebreak
\begin{equation}
8\pi^{2}\sqrt{\pi}K_{b}\left(\dfrac{\LNeck}{\SurfaceArea-\SurfaceAreaHead}\right)^{2}-\fHead/\sqrt{\SurfaceAreaHead}=0.\label{eq:spine-maturation-implicit-equation-for-spine-head}
\end{equation}
Given numerical values of $\fHead,K_{b},\SurfaceArea$ and $\LNeck$,
solving equation~(\ref{eq:spine-maturation-implicit-equation-for-spine-head})
for~$\SurfaceAreaHead$ returns all other geometrical quantities,
e.g.~$\RNeck,\RHead$ and $\SurfaceAreaNeck$. We have numerically
solved this equation for a range of values for the total surface-area~$\SurfaceArea$
and number of actin filaments in the spine-head~$N=\fHead/\fActin$.
The influences of \emph{growth}, encoded in the total surface-area
of the spine~$\SurfaceArea$, and \emph{cytoskeletal organization},
encoded in the number of actin filaments~$N$, on spine morphology
are combined in figure~\ref{fig:spine-maturation-panel}(d). Using
equation~(\ref{eq:spine-maturation-implicit-equation-for-spine-head})
and solving for the radius of the spine-neck we find radii~$\RNeck=60-93\,\mathrm{nm}$
(whereby we use estimates for the number of actin filaments $N=71$
and the typical surface-areas~$\SurfaceArea=0.5-2.00\,\mathrm{\mu m^{2}}$,
cf.~table~\ref{tab:ballpark-figures}). This range agrees quite
well with the experimentally observed ranges~$\RNeck=45-105\,\mathrm{nm}$
by~\cite{Harris1989} and~$\RNeck=50-100\,\mathrm{nm}$ by~\cite{Toennesen2014}.
From the similarity of these ranges, we infer that at least a substantial
part of the force that is exerted by the actin filaments in the spine-head
is directed towards counteracting the expansive force of the spine-neck.
Moreover, given numerical values of the total surface-area of the
spine~$\SurfaceArea$ we can solve for the number of actin filaments
required for sustaining the spine-neck. We have done this for a wide
range of surface-areas and reproduced the results in figure~\ref{fig:spine-maturation-panel}(b).
These computations show that a larger spine-head (with the same total
quantity of membrane) requires more actin filaments to sustain it.
This is in agreement with findings by~\cite{Frost2010} that show
that the number of actin filaments increases substantially with increasing
surface-area. In fact, the datapoints published in \cite{Frost2010}
match our model for $N(\SurfaceAreaHead)$ (see figure\,\ref{fig:spine-maturation-panel}(c)). We have
further used equation~(\ref{eq:spine-maturation-implicit-equation-for-spine-head})
for computing the ratios or radii~$\RHead/\RNeck$ and of volumes~$\ParameterForHead{\Volume}/\ParameterForNeck{\Volume}$,
shown in figure~\ref{fig:spine-maturation-panel}(c). We have found
that both the numerical values of~$\RHead/\RNeck$ and the upward
trend w.r.t.~$\SurfaceAreaHead$ of this metric agree well with data
published by~\cite{Toennesen2014} if we use for the number of actin
filaments $N\approx70-150$. Thus, the renders shown in figure~\ref{fig:spine-maturation-panel}(d)
appear to be in the physiologically relevant regime. Moreover, this
number for the actin filaments appears to be supported by empirical
data that shows $N\approx50-150$ (we estimated this on the basis
of data published in~\cite{Frost2010}, see Supporting Information).

\subsection{Relationship between Actin-Membrane Anchoring and Septin-Complexes
on Spine Morphology}

\begin{figure}
\centering\includegraphics{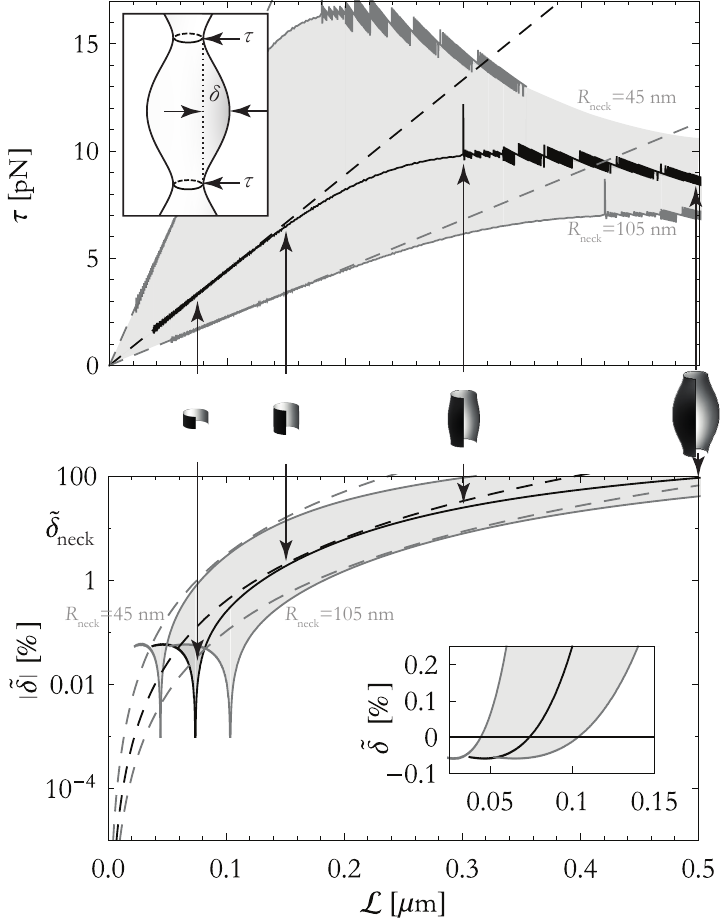}

\caption{Results of simulations that have been performed using the energy functional~(\ref{eq:CH-energy-functional})
(solid curves) and theoretical model that treats these shapes as cylinders
(dashed curves). The computations have been performed for $\protect\RNeck=45\ldots105\,\mathrm{nm}$
as indicated in the figure. Black lines corresponds to $\protect\RNeck=75\,\mathrm{nm}$.
We used $K_{b}=5\times10^{-19}\,\mathrm{J}$ for these computations~\cite{Semrau2008}.
Top panel: The line tension $\tau$ as a function of the distance
$\protect\Length$ between the line tensions. Inset shows how line
tension $\tau$ and `unduloid amplitude' $\delta$ are defined. Bottom
panel: The absolute value of the reduced `unduloid amplitude', $\left|\protect\red{\delta}\right|$.
Inset panel shows the reduced `unduloid amplitude' where it crosses
$\protect\red{\delta}=0$. Indicated is $\protect\red{\delta}_{\mathrm{neck}}=10\%$,
the approximate amplitude of variations in the width of the spine-neck,
corresponding to $\tau\approx6.5-15\,\mathrm{pN}$.\label{fig:unduloids}}
\end{figure}

We have discussed possible links between the cytoskeleton and spine
morphology and how, within our model, \emph{pushing }the spine membrane
at the location of the head effectively \emph{pulls }the spine membrane
inwards at the location of the spine-neck. Within our paradigm of
the conservation of membrane, directly applying a contractile force
$\tau$ at one or more locations along the spine-neck can achieve
the same result. This is possible due to the nature of the spine membrane,
which can be regarded as a two-dimensional fluid---that is, contracting
the spine-neck effectively channels membrane to the spine-head. Thus,
applying a line tension can aid in the transition from an immature
to a mature spine with a long, thin neck and bulbous head. Possible
candidates for such line tensions are anchoring molecules (such as
the WASP/WAVE network~\cite{Borisy2000,Takenawa2007}) septin-complexes
that form ring-like structures~\cite{Mostowy2012} and spectrin~\cite{Xu2013}.
Next, we will show that anchoring molecules or ring-like complexes
are able to apply sufficient contractile force along the spine-neck.

A line tension can be included in our models by adding a term $\tau\Circumference$
to the Canham-Helfrich free energy~(\ref{eq:CH-energy-functional}),
where $\tau$ is the line tension and $\Circumference=2\pi\RNeck$
is the circumference of the spine-neck. The line tension can be measured
thus~$\tau=-\partial\Energy/\partial\Circumference$, where $\Energy$
is the bending energy of the shape. In figure~\ref{fig:unduloids}
it can be seen that the line tensions are typically in the order of
piconewtons. As a consequence of one or a number of such line tensions
we find `unduloidal'\emph{ }spine-necks. Some representative shapes
along with the required line tension have been reproduced in figure~\ref{fig:unduloids}.
The shapes are characterized by an unduloid amplitude $\delta$
which describes the maximum deviation from the base value $\RNeck$
(we have chosen to use the relative unduloid amplitude $\red{\delta}=\delta/\RNeck$).

Although we have not been able to find publications that measure the
`unduloid amplitude' $\red{\delta}$ for spine-necks, we have calculated
this is in the order $\sim10\%$ or less (see Supplementary Information
for details). Using this value of $\red{\delta}$, we find that---if
line tensions are responsible for the typical spine-morphology---the
line tensions need to be placed at distances of $\Length\approx0.14-0.33\,\mathrm{\mu m}$,
as can be readily verified by examining figure~\ref{fig:unduloids}.
Then, computing the line tension (using numerical values $\RNeck$
and $K_{b}$ from table~\ref{tab:ballpark-figures}) corresponding
to $\Length\approx0.14-0.33\,\mathrm{\mu m}$, we find that each of
the line tensions experiences a load of $\tau\approx6.5-15\,\mathrm{pN}$.
This, too, can be verified by examining figure~\ref{fig:unduloids}.

We are aware of various candidates for anchoring membrane to the cytoskeleton,
such as L-selectin, $\beta_{2}$ integrins and CD45. The literature
reports that these three candidates have rupture forces respectively
$25-45\,\mathrm{pN}$, $60-120\,\mathrm{pN}$ and $35-85\,\mathrm{pN}$~\cite{Shao1999}.
Even the lowest values of these three ranges is almost double our
highest estimate for the required line tension. Therefore, it is safe
to conclude that anchoring molecules can withstand the mechanical
forces that are required in order to constrain the spine-neck to $\RNeck=45-105\,\mathrm{nm}$.

Since spine-necks typically have lengths of $0.2-2\,\mathrm{\mu m}$
(table~\ref{tab:ballpark-figures}, see Supporting Information),
we find that the number of line tensions that needs to be placed is
$1-14$ (whereby we used the aforementioned distance between the line
tensions $\Length\approx0.14-0.33\,\mathrm{\mu m}$). However---as
far as we can determine---literature does not mention these concentrations
of anchoring molecules across all spine-necks. Although ring-like
septin-complexes are found consistently along spine-necks~\cite{Ewers2014,Mostowy2012,Tada2007},
they are only reported to be positioned at the \emph{base }of the
spine and \emph{not} along the full length of the spine-neck. Our
models predict that it is required to place line tensions along the
full length of the spine-neck in order to constrain it, and therefore
we can refute septin-complexes as being solely responsible for constraining
the long, thin spine-necks. Moreover, the assembly of septins into
ring-like structures has an associated time-scale in the order of
minutes~\cite{Kinoshita2002}. Hence, we find that cytoskeletal remodeling---which
can performed on the time-scale of fractions of a second~\cite{Mogilner2003}---is
much more rapid than positioning these constriction proteins.

\section{Conclusions}

We study the physical mechanisms that determine the morphology of dendritic spines. In particular, we investigate the ability of the actin cytoskeleton to change the size and shape of spines. We find that the most striking primary features of spine growth and spine morphology can be straightforwardly understood as a consequence of the trade-off between the elastic properties of the spine membrane and the forces actively generated by the actin cytoskeleton. Specifically, we show that the initiation and formation of dendritic filopodia may be rationalized on the basis of the protrusive forces of the actin cytoskeleton. Using realistic estimates for the number of actin filaments involved, we find that the dimensions of the filopodia in our models agrees well with the observed dimensions of newly formed protrusions in the developing neuron.

We have also studied spine maturation, the process characterized by
a morphological transition from a filopodium or thin spine to the
mature mushroom-like spine. Using models based on the coupling
between the actin cytoskeleton and the spine membrane, we find that the
combined dynamics of branched actin and aligned actin inherently results
in a mushroom-like morphology. Finally, we have discussed several candidate factors that might aid in the stabilization of the long, thin spine-neck and the bulbous spine-head. Our predictions for various morphological quantities, such as the neck radius $\RNeck$
and the ratios $\RHead/\RNeck$ and $\ParameterForHead{\Volume}/\ParameterForNeck{\Volume}$, compare well with experimental data~\cite{Toennesen2014,Harris1989}. Furthermore, the suggested important dual roles of branched and aligned organizations suggest novel experiments analyzing (possibly, even, altering) the localization of proteins like Arp2/3 and septin to the spine head and neck. Summarizing, our model suggests that actin organization is autonomously capable of controlling the shape changes of dendritic spines, providing the forces and geometry support for both the initial filopodial stage and the mature mushroom-like shape.

\section{Supporting Information}

\subsection{Estimate for the Number of Actin Filaments\label{sub:SI-number-of-actin-filaments-from-korobova}}

We counted the number of actin filaments as $20$ on $\sim20\%$ of
the surface-area resulting in $\sim100$ filaments for one entire
spine-head as published by~\cite{Korobova2010}. Then, noting that
\emph{on the average} the filaments are not oriented perpendicular
to the membrane--but rather at an angle $\pi/4$, we find the \emph{effective}
number of actin filaments to be $\sim100\cdot\cos(\pi/4)\approx71$.
This number falls within the range for the number of polymerizing
filaments~$N=50-150$ we derived from data published in~\cite{Frost2010}
(\cite{Frost2010} has published the density of non-stationary actin
molecules, which we integrated over the surface area to obtain a measure
for the number of polymerizing filaments).

\subsection{Standard Deviation in Spine-Neck Width\label{sub:SI-variation-in-spine-neck-widths}}

We measured the width of the spine-neck of images by~\cite{Toennesen2014}
by fitting the intensity of the profile with Gaussian distributions
along the axis of the spine-neck. We asserted that the standard deviation
of these Gaussians is a measure for the width of the spine-neck. Then,
we computed the relative variation in these widths. Using this method,
the relative variation in the width of the spine-neck was found to
be $13.5\%$.

\subsection{Shape Equations}

Taking the first variation of the Canham-Helfrich energy functional
(\ref{eq:CH-energy-functional}), and insisting that the first variation
$\delta\Energy$ is zero under all possible infinitesimal perturbations
results in a differential equation that describes stationary shapes
$\{r(s),\psi(s)\}$.\footnote{The stationary shapes include shapes corresponding to an energetic
minimum, an energetic maximum or a saddle point in the energy functional.
A seminal paper by \cite{OuYong1989} describes the higher-order variations,
from which we can infer the class of stationary point. We will not
discuss this technical difficulty in this publication, although we
have used numerical perturbative methods to determine which shapes
correspond to energetic minima.} This differential equation, that we shall henceforth call the \emph{shape
equation}, is \cite{Juelicher1994,Derenyi2002,Mathews2012}\nopagebreak
\begin{multline}
\psi^{(3)}=-\dfrac{1}{2}(\psi')^{3}-\dfrac{2\cos\psi}{r}\psi''+\dfrac{3\sin\psi}{2r}(\psi')^{2}+\dfrac{\bar{\sigma}}{r}\sin\psi\\
+\dfrac{3(\cos\psi)^{2}-1}{2r^{2}}\psi'+\bar{\sigma}\psi'-\dfrac{(\cos\psi)^{2}+1}{2r^{3}}\sin\psi,\label{eq:shape-equation}
\end{multline}
where we have dropped the $s$--dependencies and $\bar{\sigma}\equiv\sigma/K_{b}$.
Most publications that we have consulted make reference to second-order
shape equations~\cite{Juelicher1994}, but--in accordance with~\cite{Derenyi2002}--we
find the third-order shape equation~(\ref{eq:shape-equation}) to
be numerically substantially more stable. The second-order shape equations,
e.g.~found by taking the first integral of (\ref{eq:shape-equation}),
is used to find boundary conditions for $\psi''$. This equation is~\cite{Derenyi2002}\nopagebreak
\begin{multline}
\psi''\cos\psi=-\dfrac{1}{2}(\psi')^{2}\sin\psi-\dfrac{(\cos\psi)^{2}}{r}\psi'+\\
\dfrac{(\cos\psi)^{2}+1}{2r^{2}}\sin\psi+\bar{\sigma}\sin\psi-\dfrac{\bar{f}}{r},\label{eq:shape-equation-second-order}
\end{multline}
where $\bar{f}\equiv f/K_{b}$. Although the point force $f$ does
not show up in the shape equation (\ref{eq:shape-equation}), it does
enter in the determination of the correct boundary conditions through
(\ref{eq:shape-equation-second-order}).

We use a shooting-and-matching algorithm (for more information on
this numerical technique, we refer to \cite{Heath2002})  whereby
$\psi'(0),f$, $\sigma$ are used as shooting variables.

\bibliographystyle{abbrv}
\bibliography{GraduateThesis}


\end{document}